%% file: main.tex
\documentclass[conference]{IEEEtran}
\IEEEoverridecommandlockouts
\usepackage{cite}
\usepackage{amsmath,amssymb,amsfonts}
\usepackage{algorithm}
\usepackage{algpseudocode}
\usepackage{graphicx}
\usepackage{textcomp}
\usepackage{xcolor}
\usepackage{multirow}
\usepackage{booktabs}
\usepackage{setspace}
\usepackage{comment}
\usepackage{enumitem}
\usepackage{setspace}
\usepackage{caption}
\usepackage{subcaption}
\newlist{inlineroman}{enumerate*}{1}
\setlist[inlineroman]{itemjoin*={{ }},afterlabel=~,label=\roman*.}

\renewcommand{\baselinestretch}{0.97}

\newcommand{\inlinerom}[1]{
\begin{inlineroman}
#1
\end{inlineroman}
}
\algblock{Input}{EndInput}
\algnotext{EndInput}
\algblock{Ensure}{EndEnsure}
\algnotext{EndEnsure}
\algblock{Output}{EndOutput}
\algnotext{EndOutput}

\def\BibTeX{{\rm B\kern-.05em{\sc i\kern-.025em b}\kern-.08em
    T\kern-.1667em\lower.7ex\hbox{E}\kern-.125emX}}

\begin{document}
\captionsetup[table]{labelfont=bf,textfont=normalfont}

\title{
Accelerating Dynamic Image Graph Construction on FPGA for Vision GNNs
}



\author{
\IEEEauthorblockN{Anvitha Ramachandran, Dhruv Parikh, Viktor Prasanna}
\IEEEauthorblockA{
   University of Southern California, Los Angeles, California, USA \\
    alramach@usc.edu, dhruvash@usc.edu, prasanna@usc.edu}
}

\maketitle

\begin{abstract}
Vision Graph Neural Networks (Vision GNNs, or ViGs) represent images as unstructured graphs, achieving state-of-the-art performance in computer vision tasks such as image classification, object detection, and instance segmentation. Dynamic Image Graph Construction (DIGC) builds image graphs by connecting patches (nodes) based on feature similarity, and is dynamically repeated in each ViG layer following GNN-based patch (node) feature updates. However, DIGC constitutes over $50\%$ of end-to-end ViG inference latency, rising to $95\%$ at high image resolutions, making it the dominant computational bottleneck. While hardware acceleration holds promise, prior works primarily optimize graph construction algorithmically, often compromising DIGC’s flexibility, accuracy, or generality.

To address these limitations, we propose a streaming, deeply pipelined FPGA accelerator for DIGC, featuring on-chip buffers that process input features in small, uniform blocks. Our design minimizes external memory traffic via localized computation and performs efficient parallel sorting with local merge-sort and global k-way merging directly on streaming input blocks via heap insertion. This modular architecture scales seamlessly across image resolutions, ViG layer types, and model sizes and variants, and supports DIGC across diverse ViG-based vision backbones. The design achieves high clock frequencies post place-and-route due to the statically configured parallelism minimizing critical path delay and delivers up to $16.6\times$ and $6.8\times$ speedups over optimized CPU and GPU DIGC baselines.

\end{abstract}



\input{0-intro}

\input{1-related}
\input{2-prelims}
\input{3-method}
\input{4-exp}

\input{5-concl}

\bibliographystyle{IEEEtran}
\bibliography{IEEEabrv,citation}

\end{document}

%% file: 0-intro.tex
\section{Introduction}
\label{sec:intro}

\noindent 
The landscape of computer vision has been shaped by a succession of architectural innovations, beginning with Convolutional Neural Networks (CNNs) \cite{CNNs}, advancing through Multi-Layer Perceptron models \cite{MLP-Mixer}, and more recently, Vision Transformers (ViTs) \cite{VisionTransformers}. CNNs excel at local feature extraction \cite{AlexNet}, MLPs offer flexible global modeling \cite{ResMLP}, and ViTs  leverage self-attention to capture long-range dependencies \cite{SwinTransformer}. Vision Graph Neural Networks (ViGs) have emerged as an alternative, representing images as general-purpose graphs where nodes correspond to image patches and edges encode relationships \cite{VisionGNN}. Unlike CNNs, ViTs and MLPs, which are bound by structured representation, ViGs dynamically construct graphs based on feature similarity. This flexible representation enables Vision GNNs to capture both local and global context, outperforming CNNs, ViTs and MLPs in tasks that require efficient visual reasoning \cite{ViHGNN}.

\begin{figure*}[t]
  \centering
  \begin{subfigure}[t]{0.48\textwidth}
    \centering
    \includegraphics[width=\textwidth]{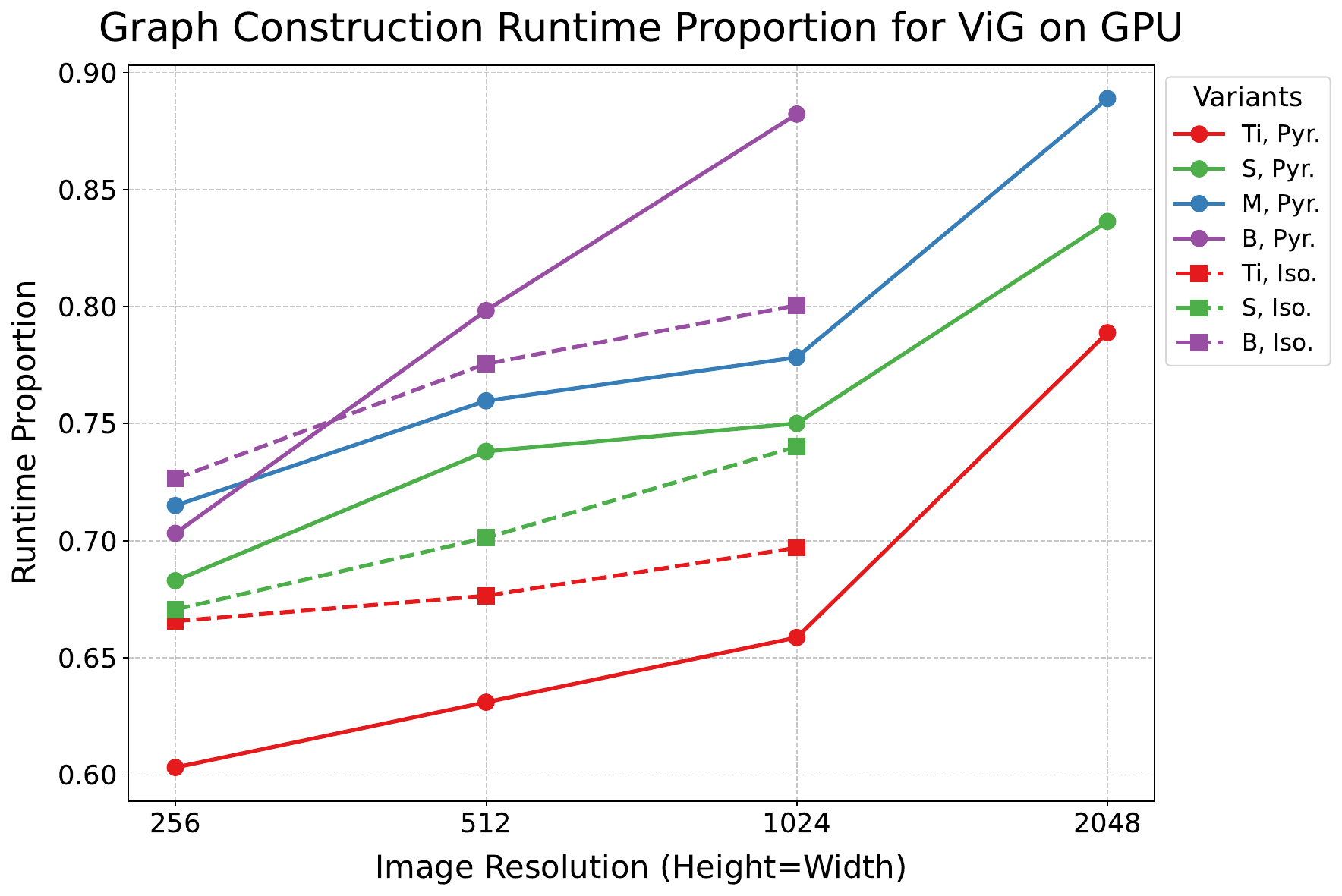}
    \caption{GPU results. Lines not extending towards 2048 were due to out-of-memory errors.}
    \label{fig:figGPU}
  \end{subfigure}
  \hfill
  \begin{subfigure}[t]{0.48\textwidth}
    \centering
    \includegraphics[width=\textwidth]{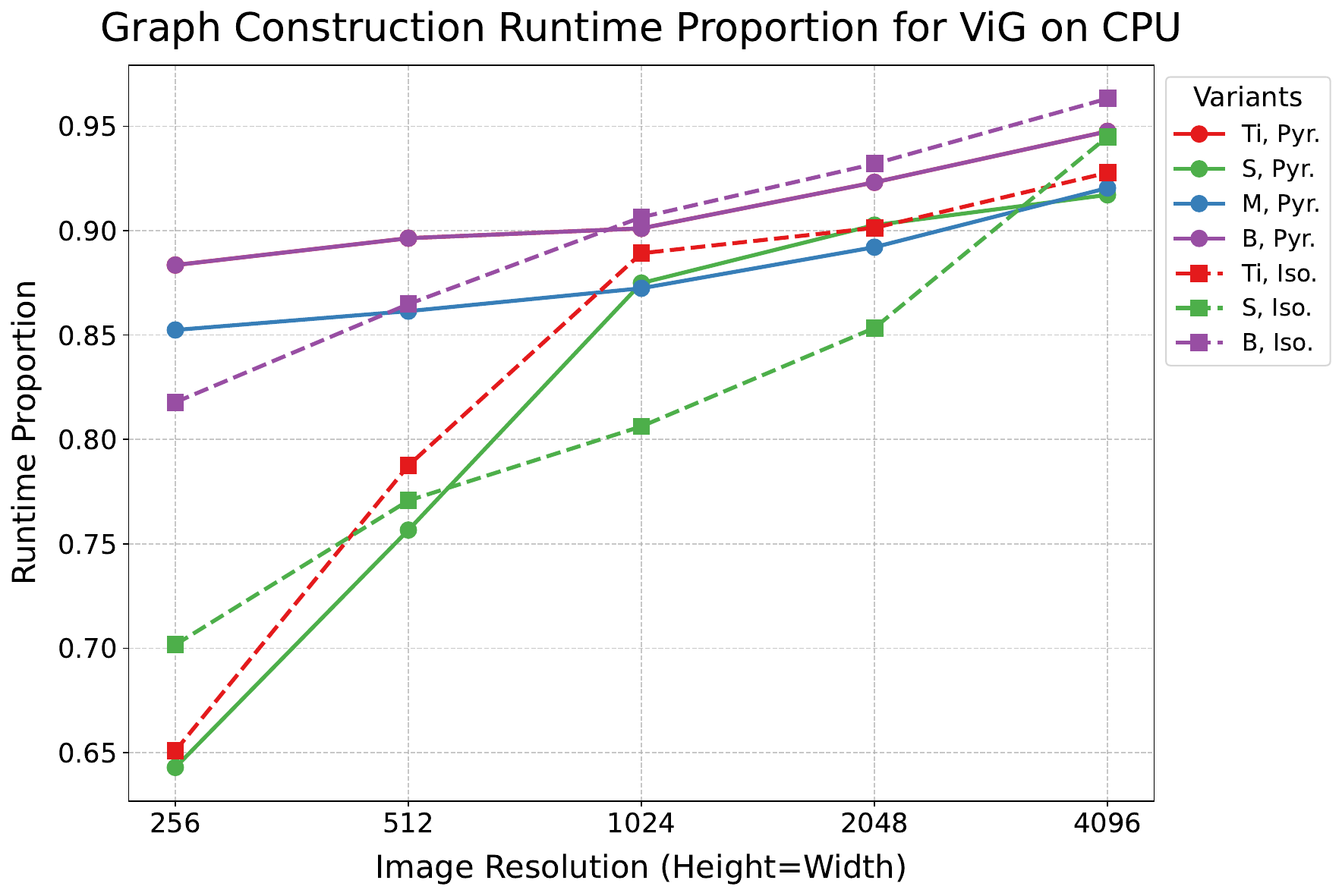}
    \caption{CPU results. Same model variants as in (a).}
    \label{fig:figCPU}
  \end{subfigure}
  \caption{Percentage of end-to-end inference latency across different ViG model sizes (Ti: Tiny, S: Small, M: Medium, B: Base) and architectures (Pyr.: Pyramidal, Iso.: Isotropic) as a function of input image resolution. Batch size was fixed at $1$.}
  \label{fig:vig_profiles}
\end{figure*}

ViG models have been widely adopted across core computer vision tasks including image classification~\cite{wang2023two}, object detection~\cite{GeoGraph}, and instance segmentation~\cite{Chaidos}, as well as diverse applied domains. In medical imaging, they improve tumor detection, brain disorder diagnosis, and denoising for MRI, CT, and EEG data~\cite{BrainTumor, TumorCell, medicalgnn, eeggraph}; in video analysis, they enable real-time anomaly detection, event recognition, and segmentation~\cite{VideoBased, VideoSeg, gnnvideo}; in materials science, they support property prediction, crystal analysis, and virtual sensing~\cite{yang2024svgacrack, lapenna2025vision, matgnn}; and in industrial applications, they facilitate CAD model classification, structural monitoring, and part retrieval~\cite{cadgnn, shmgnn}. Real-time inference is especially critical in healthcare, video diagnostics, and scientific workloads. However, the Dynamic Image Graph Construction (DIGC) operation in ViGs, which partitions input images into patches and establishes edges based on spatial or feature similarity, introduces substantial computational overhead. Further, as the resolution of input image grows, the number of graph nodes and potential edges increases quadratically, amplifying the computational and memory demands of DIGC \cite{ClusterViG}.  The impact of this bottleneck varies based on the ViG architecture. Isotropic ViG variants process all nodes and edges uniformly across every network layer and spatial scale. Pyramidal ViG variants adopt a hierarchical approach, where node features are pooled and the graph is reconstructed on progressively lower-resolution feature maps, resulting in improved accuracy. In these models, graph construction can consume up to 90\% of inference time at high resolutions, severely limiting scalability. Our profiling, illustrated in Figures \ref{fig:figGPU} and \ref{fig:figCPU}, reveals that graph construction is the dominant computational bottleneck in end-to-end ViG model inference. For larger image resolutions, the runtime proportion of the graph construction stage becomes larger across both isotropic and pyramidal variants. 

While CPUs and GPUs have been the standard platforms for ViG inference, their general-purpose architectures struggle with the fine-grained, irregular memory access patterns intrinsic to DIGC workloads. Graph construction involves pairwise distance computation, sorting, and neighbor selection—operations that require non-uniform memory accesses and high data movement. Further, conventional CPU and GPU implementations execute these stages sequentially, offloading intermediate results to external memory, significantly increasing memory traffic which leads to severe performance degradation, especially at high input resolutions. In contrast, FPGAs offer the flexibility to implement tightly pipelined, on-chip dataflows that minimize off-chip memory traffic and efficiently handle the heterogeneity of graph construction. In this work, we propose an end-to-end FPGA accelerator for efficient DIGC computation on FPGA. Our contributions are as follows:

\begin{itemize}
    \item We propose a fully streaming, deeply pipelined FPGA architecture for Dynamic Image Graph Construction (DIGC) in Vision Graph Neural Networks (ViGs).
    \item Our accelerator features a scalable design using data partitioning, distance computation and sorting modules, supporting various image sizes and ViG variants without hardware changes.
    \item The modular architecture enables efficient placement and routing, supporting a high post place-and-route clock frequency and efficient resource utilization on the Xilinx Alveo U280 FPGA.
    \item We validate the fixed-configuration design on multiple ViG models and resolutions up to 2048×2048, showing robust performance improvements even where GPU baselines fail due to out-of-memory error.
    \item Our HLS-based design, deployed on Xilinx Alveo U280, achieves up to $16.6\times$ speedup over CPUs and $6.8\times$ over GPUs on DIGC for a single image, resulting in $2.1\times$--$4.6\times$ end-to-end inference latency acceleration across ViG variants.    
    \item The proposed accelerator scales efficiently across input image resolutions and ViG layer types and variants.
\end{itemize}

%% file: 1-related.tex
\section{Related Work}
\label{sec:related}

Recent advances in Vision Graph Neural Networks (ViGs) demonstrate strong performance by dynamically constructing graphs over image patches to capture spatial dependencies. Pioneering works such as ViG~\cite{VisionGNN}, ViHGNN~\cite{ViHGNN}, PVG~\cite{PVG}, and ClusterViG~\cite{ClusterViG} introduce various strategies for adaptive graph construction, balancing local and global context. Extensions like GreedyViG~\cite{GreedyVIG}, MobileViG~\cite{MobileViG, MobileViG2}, and WiGNet~\cite{WiGNet} focus on efficiency through sparse or structured graph representations, while DVHGNN~\cite{DVHGNN} and VSViG~\cite{VideoBased} address receptive field expansion and video applications. ViGs have further shown domain versatility across medical imaging~\cite{BrainTumor,TumorCell,lapenna2025vision} and scientific datasets~\cite{ClusterViG,PVG}.

Existing acceleration efforts primarily target general GNN workloads using algorithmic sparsification~\cite{GreedyVIG,ClusterViG}, quantization~\cite{MEGA}, and structured graph construction~\cite{WiGNet}. Hardware accelerators span CPU/GPU frameworks~\cite{Graphite,FTC-GNN}, FPGA designs~\cite{GCVTurbo,GraphAGILE, MaGNAS, EvGNN}, and in-memory architectures~\cite{PIM-GCN,Lift}. However, none directly address ViG-specific bottlenecks. Our work introduces the first ViG-specialized FPGA accelerator that targets the dominant Dynamic Image Graph Construction (DIGC) phase, combining ViG-specific optimizations with a fully pipelined, memory-efficient hardware design.

%% file: 2-prelims.tex
\section{Preliminaries}
\label{sec: prelim}


In this section, we provide a formal definition of GNNs, describe the architecture of ViG models, and outline the methodology for constructing image graphs. Finally, we formally define the problem of Dynamic Image Graph Construction (DIGC).

\subsection{Graphs}
A graph ($\mathcal{G}(\mathcal{V},\mathcal{E})$ is defined as a set of vertices $\mathcal{V}$, and a set of edges $\mathcal{E}$ connecting two of the vertices. Directed edges can be defined as $\mathcal{E} = \{(j, i)\ |\ j, i \in \mathcal{V}\}$ such that ordered pairs specify the direction. Each vertex $i \in \mathcal{V}$ has a feature vector $x_i = \mathbb{R}^D$. The feature matrix $\boldsymbol{X} \in \mathbb{R}^{N\times D}$ contains feature vectors of all the vertices such that the node set $\mathcal{V} = \{0, 1, ..., N-1\}$ corresponds to feature vector $x_{i\in\mathcal{V}} = \boldsymbol{X}(i,:)$.

\subsection{Graph Neural Networks}
Given an input graph $\mathcal{G}(\mathcal{V},\mathcal{E})$ with a node feature matrix $\boldsymbol{X}$,  an $L$ layered GNN can
be described using the message passing framework as follows \cite{GNN_MP}:

\begin{equation}
\label{eq:gnn}
\mathbf{x}_i^{(l)} = \Psi^{(l)}\left( \mathbf{x}_i^{(l-1)},\ \oplus_{j \in \mathcal{N}(i)} \Phi^{(l)}\left( \mathbf{x}_i^{(l-1)}, \mathbf{x}_j^{(l-1)} \right) \right)
\end{equation}

Message passing in eq. \ref{eq:gnn} is applied $\forall i \in V,  \forall l \in \{1, 2, ... L\}$, and the input to the $l^\text{th}$ layer of the GNN is the output of the previous layer, $\boldsymbol{X}^{(l-1)} \in \mathbb{R}^{N\times D^{(l-1)}}$. For $l=1$, we have $\boldsymbol{X}^{(0)} = \boldsymbol{X}, D^{(0)} = D$. The set of neighbors for node $i$ is denoted as $\mathcal{N}(i) = \{j\;\vert \; (j, i) \in \mathcal{E}\}$. Note that $\oplus$ is an aggregation operation such as sum, max or mean, and $\Psi^{(l)}, \Phi^{(l)}$ are learnable functions such as MLPs.

The feature vector of every node $i\in\mathcal{V}$ is updated in each layer $l$ of the GNN in three core operations: \inlinerom{\item \textit{Message Creation}. For each node $j$ connected to node $i$ via edge $(j,i)$, a message $\mathbf{m}_{j,i}^{(l)} = \Phi^{(l)}(\mathbf{x}_i^{(l-1)}, \mathbf{x}_j^{(l-1)})$ is created. \item \textit{Aggregation}. The messages are aggregated over $\mathcal{N}(i)$ as $\mathbf{a}_i^{(l)}= \oplus_{j \in \mathcal{N}(i)} \mathbf{m}_{j,i}^{(l)}$. \item \textit{Update}. The feature vector of node $i$ is updated as $\mathbf{x}_{i}^{(l)} = \Psi^{(l)}(\mathbf{x}_i^{(l-1)}, \mathbf{a}_i^{(l)})$.}

\subsection{Vision Graph Neural Networks (ViG)}
In Vision Graph Neural Networks (ViG), the input image tensor of dimensions $H \times W \times C$ is partitioned into $N = \frac{H \times W}{P^2}$ non-overlapping patches of size P (tokens), each embedded into a $D$-dimensional feature vector via a convolutional stem, yielding a node feature matrix $X \in \mathbb{R}^{N \times D}$. 
A co-node feature matrix $ Y \in \mathbb{R}^{M \times D} $ is constructed by applying spatial pooling to the original feature tensor $ X \in \mathbb{R}^{\frac{H}{P} \times \frac{W}{P} \times C} $, reducing each spatial dimension by a factor $ r $ to form a smaller tensor of shape $ \frac{H}{P_r} \times \frac{W}{P_r} \times C $. This is then reshaped to $Y \in \mathbb{R}^{\frac{N}{r^2} \times D}$, where $r^2$ is the total spatial reduction factor. This pooling is done in early pyramid ViG layers with large resolution to reduce distance matrix computations by approximately $r^2$. In contrast, isotropic ViG models maintain a steady small resolution without this reduction. Graph connectivity is controlled by $k$, the number of nearest neighbors per node.

\begin{figure*}
\centering
\includegraphics[width=0.9\textwidth]{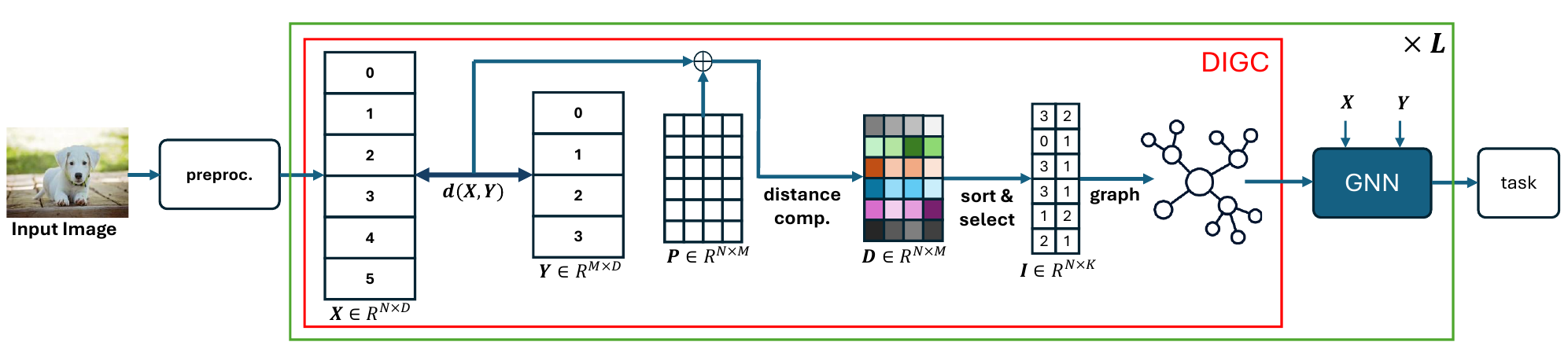}
\footnotesize{\caption{Overview of the Dynamic Image Graph Construction (DIGC) pipeline in Vision Graph Neural Networks (ViGs). The figure illustrates the input image patches ($\mathbf{X}$), co-node features ($\mathbf{Y}$), positional embeddings ($\mathbf{P}$), and the computation flow through distance computation, sorting, and graph neighbor selection modules, highlighting the data dimensions and integration with the graph neural network (GNN) task.
\label{fig:vig_explained}}}
\end{figure*}
\subsection{Dynamic Image Graph Construction (DIGC)}
\label{sec:DIGC}
In ViGs, a central challenge is the dynamic construction of an image graph. Given an input image, the goal is to partition it into patches and construct a graph where patches representing nodes and edges are formed based on feature similarities based on the learned feature vectors. DIGC enables ViG models to flexibly model complex visual dependencies \cite{VisionGNN}. The ViG model uses dilated k-nearest neighbor (KNN) graph construction, in which each node first identifies its $k \times d$ nearest neighbors, then selects every $d$-th neighbor from this sorted list to form its $k$ graph connections. This dilation strategy enlarges the receptive field without loss of resolution \cite{DVHGNN}. We define the serial algorithm that motivates our methods for DIGC in Algorithm \ref{alg:ser}.
\begin{algorithm}[htp]
\caption{Serial DIGC}\label{alg:ser}
\footnotesize{
\begin{algorithmic}[1]
  \Input\;$X \in \mathbb{R}^{N\times D}$, Feature vectors of vertices,
  $Y \in \mathbb{R}^{M\times D}$, Feature vectors of co-vertices,
  $P \in \mathbb{R}^{N\times M}$ Relative positional embeddings, $k \in \mathbb{N}$, the set of natural numbers,
  $d \in \mathbb{N}$ Dilation factor.
  \EndInput
  \Output\;$I \in \{0,1,...,M-1\}^{N\times k}$ Nearest neighbor indices for each vertex
  \EndOutput

\State $\text{inner}_{XY} \leftarrow -2XY^T$
\State $\text{square}_X \leftarrow \left[ \sum_{j=0}^{N-1} X_{ij}^2 \;\middle|\; i = 0, \ldots, M-1 \right]$
\State $\text{square}_Y \leftarrow \left[ \sum_{i=0}^{M-1} Y_{ij}^2 \;\middle|\; j = 0, \ldots, N-1 \right]$
\State $D_{XY} \leftarrow \text{inner}_{XY} + \text{square}_X \mathbf{1}_N^T + \mathbf{1}_M \text{square}_Y^T$
\State $D_{XY} \leftarrow D_{XY} + P$

\State $\text{sorted}_{\text{index}}, \text{sorted}_D \leftarrow \text{sort}(D_{XY})$ \Comment{Ascending order}
\State $I^\prime \leftarrow \text{sorted}_{\text{index}}[:][:kd]$ \Comment{Sliced kd top neighbors}
\State $I \leftarrow I^\prime[:][::d]$ \Comment{Every $d^{th}$ element} 
\end{algorithmic}}
\end{algorithm}


%% file: 3-method.tex
\section{Methodology}
\label{sec: methods}

We introduce the architecture and implementation of our FPGA accelerator for DIGC. In section \ref{sec:Data}, we define the data layout and format. In section \ref{sec:Overview}, we describe the structure of our modular accelerator for graph construction: our Distance Computing Module (DCM), Local Sorting Module (LSM), Global Merging Module (GMM), and Neighbor Selection Module (NSM). In section \ref{sec:ComputeFlow}, we discuss the compute flow of our accelerator, ending briefly with a performance model to estimate total cycles to compute DIGC.
\subsection{Data Layout}
\label{sec:Data}
The node feature matrix \( X \in \mathbb{R}^{N \times D} \), co-node matrix \( Y^T \in \mathbb{R}^{D \times M} \), and positional embeddings \(P \in \mathbb{R}^{N\times M}\) are stored in external DDR4 memory due to their large size, particularly for larger image resolutions. DDR4 provides sufficient bandwidth and storage for large-scale ViG models and high-resolution data. Feature data is laid out in row-major order and block-wise tiling across the rows and columns of $X, Y$, $D$ and $P$ is aligned to the memory bus width for efficient burst transfers. 

\subsection{Accelerator Overview}
\label{sec:Overview}
\begin{figure}
\centering
\includegraphics[width=0.48\textwidth]{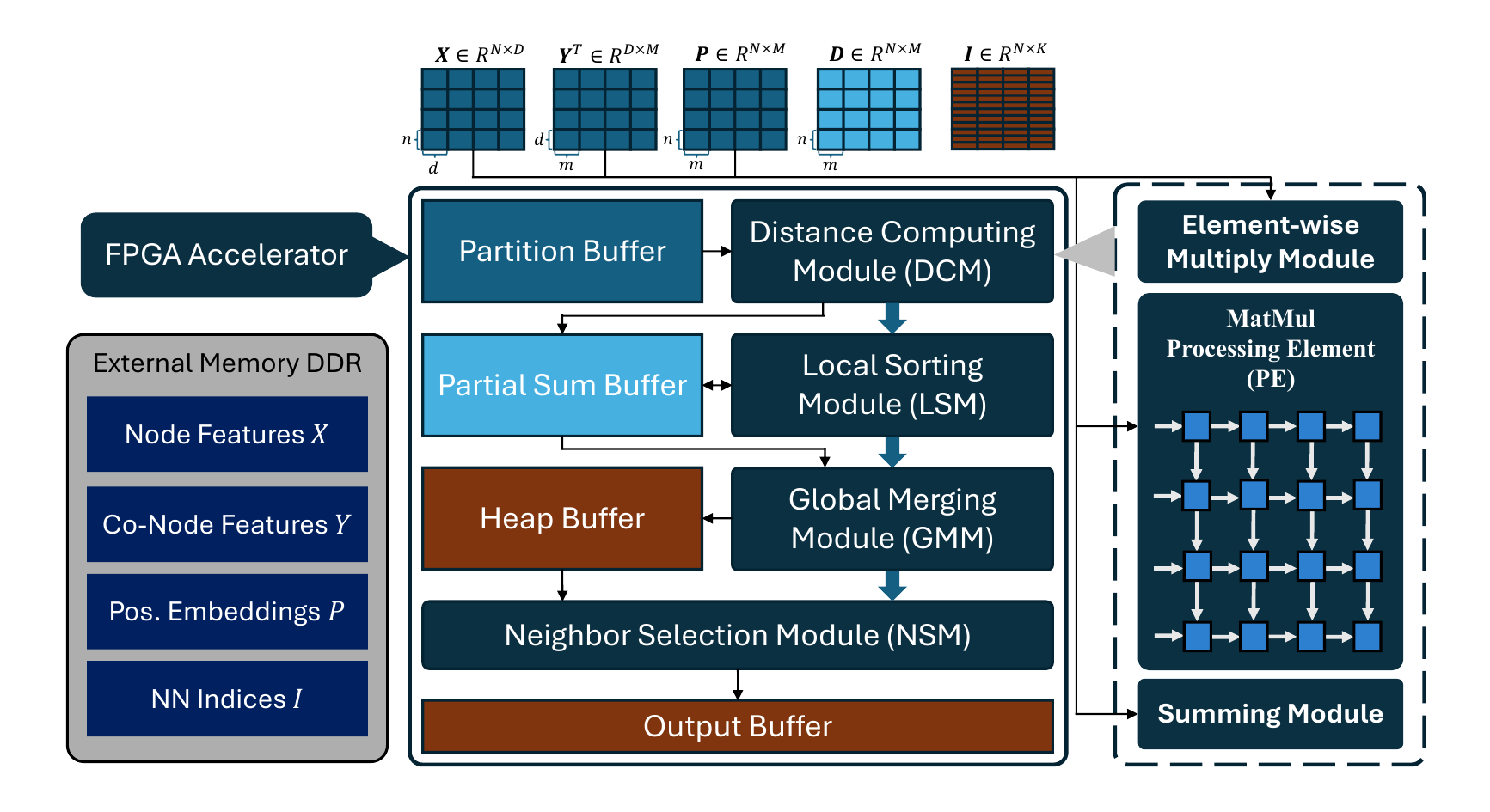}
\footnotesize
{\caption{Block diagram of the FPGA accelerator for DIGC, illustrating external memory organization, on-chip buffers, and key modules (DCM, LSM, GMM, NSM), along with data dimensions, flow, and 2D PE mesh partitioning.}
\label{fig:overview}}
\end{figure}

Our FPGA-based accelerator targets efficient image graph construction (Section~\ref{sec:DIGC}) and contains four main modules: Distance Computation (DCM), Local Sorting (LSM), Global Merging (GMM), and Neighbor Selection (NSM), as shown in Figure~\ref{fig:overview}. On-chip buffers store intermediate distances (Partial Sum Buffer, PSB), locally sorted partitions (Heap Buffer, HB), and final neighbor indices (Output Buffer, OB).  The accelerator employs a block-wise matrix multiplication PE mesh, an element-wise multiplication PE mesh, and an accumulation PE mesh. The number of row and column partitions, $P_{\text{row}}$ and $P_{\text{col}}$, control parallelism across input nodes and co-nodes, while operations along the feature dimension $D$ are parallelized with an unroll factor $P_{\text{vec}}$ for fine-grained SIMD execution. After distance computation, each PE outputs distances for its assigned partition, which are staged in local on-chip buffers and sent to specialized sorting PEs. Local sorts feed into a global merge implemented via a distributed k-way min-heap, as shown in Figure \ref{fig:lsm_gmm}.

\subsection{Compute Flow}
\label{sec:ComputeFlow}
\begin{figure}
\centering
\includegraphics[width=\columnwidth]{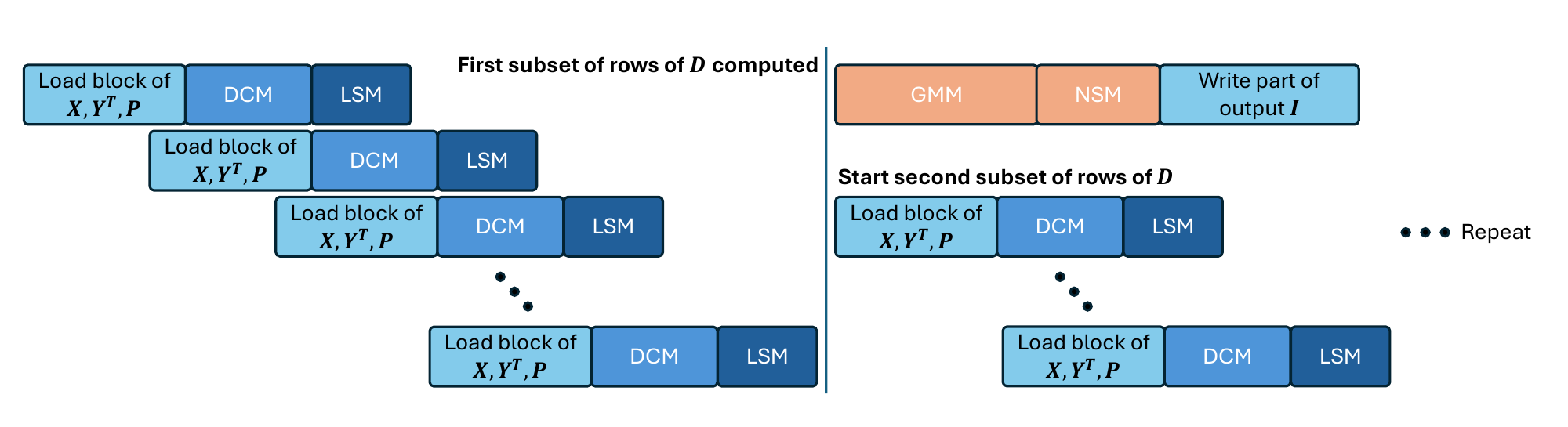}
\caption{Pipeline of the FPGA accelerator for DIGC, showing input feature loading ($\mathbf{X}$, $\mathbf{Y}^T$, $\mathbf{P}$), key modules (DCM, LSM, GMM, NSM), and the streaming, partitioned dataflow for parallel processing.}
\label{fig:pipeline}
\end{figure}

This section details the end-to-end compute flow of our FPGA-based accelerator for DIGC, depicted in Figure \ref{fig:pipeline}. The architecture is optimized to accelerate graph construction within each individual layer of the ViG model.

\subsubsection{Distance Computation Module (DCM)}

The Distance Computation Module (DCM), as shown in Algorithm \ref{alg:DCM} calculates the squared Euclidean distance between node and co-node feature vectors. The DCM is architected as a two-dimensional mesh of Processing Elements (PEs), parameterized by $P_{\text{row}}$ (parallelism over nodes) and $P_{\text{col}}$ (parallelism over co-nodes), forming a $P_{\text{row}} \times P_{\text{col}}$ grid. Each PE in the mesh is assigned a unique $(i, j)$ coordinate and is responsible for computing the distance between node $i$ and co-node $j$ within a partition. The DCM computes the squared Euclidean distance between node feature vector $\mathbf{x}_i$ and co-node feature vector $\mathbf{y}_j$ as $\text{dist}_{i,j} = \|\mathbf{x}_i - \mathbf{y}_j\|^2 = \|\mathbf{x}_i\|^2 - 2\langle \mathbf{x}_i, \mathbf{y}_j \rangle + \|\mathbf{y}_j\|^2$. All intermediate results are stored in intermediary buffers to be computed and stored in the Partial Sum Buffer, enabling pipelined accumulation and streaming dataflow. 

\begin{algorithm}
\footnotesize{
\begin{algorithmic}[1]
\caption{PE Allocation and Per-PE Distance Computation in the DCM}
\label{alg:DCM}
\Require Partitioned on-chip buffers for node features $X \in \mathbb{R}^{n \times D}$ and co-node features $Y^T \in \mathbb{R}^{m \times D}$; partition indices: \texttt{row\_base}, \texttt{col\_base}; PE mesh dimensions: $P_{\text{row}}$, $P_{\text{col}}$
\Statex
\Comment{PE Allocation}
\For{$r = 0$ to $P_{\text{row}} - 1$} \Comment{Parallel over PE mesh rows}
    \For{$c = 0$ to $P_{\text{col}} - 1$} \Comment{Parallel over PE mesh columns}
        \State $i \gets \texttt{row\_base} + r$ \Comment{Node index for this PE}
        \State $j \gets \texttt{col\_base} + c$ \Comment{Co-node index for this PE}
        \State Allocate $\text{PE}_{r,c}$ for pair $(i, j)$
    \EndFor
\EndFor

\ForAll{$\text{PE}_{r,c}$ \textbf{in parallel}}\Comment{Per-PE Computation}
    \State $\mathbf{x}_i \gets X[i, :]$ \Comment{From $r$-th bank}
    \State $\mathbf{y}_j \gets Y^T[j, :]$ \Comment{From $c$-th bank}
    \State $x\_sq \gets 0$; $y\_sq \gets 0$; $xy\_sum \gets 0$
    \For{$d = 0$ to $D - 1$}
        \State $x\_val \gets x_{i,d}$
        \State $y\_val \gets y_{j,d}$
        \State $x\_sq \gets x\_sq + x\_val \times x\_val$
        \State $y\_sq \gets y\_sq + y\_val \times y\_val$
        \State $xy\_sum \gets xy\_sum + x\_val \times y\_val$
    \EndFor
    \State $\text{dist}_{i,j} \gets x\_sq + y\_sq - 2 \times xy\_sum$
\EndFor
\end{algorithmic}}
\end{algorithm}

\subsubsection{Local Sorting Module (LSM)}

\begin{figure}
\footnotesize{
\centering
\includegraphics[width=\columnwidth]{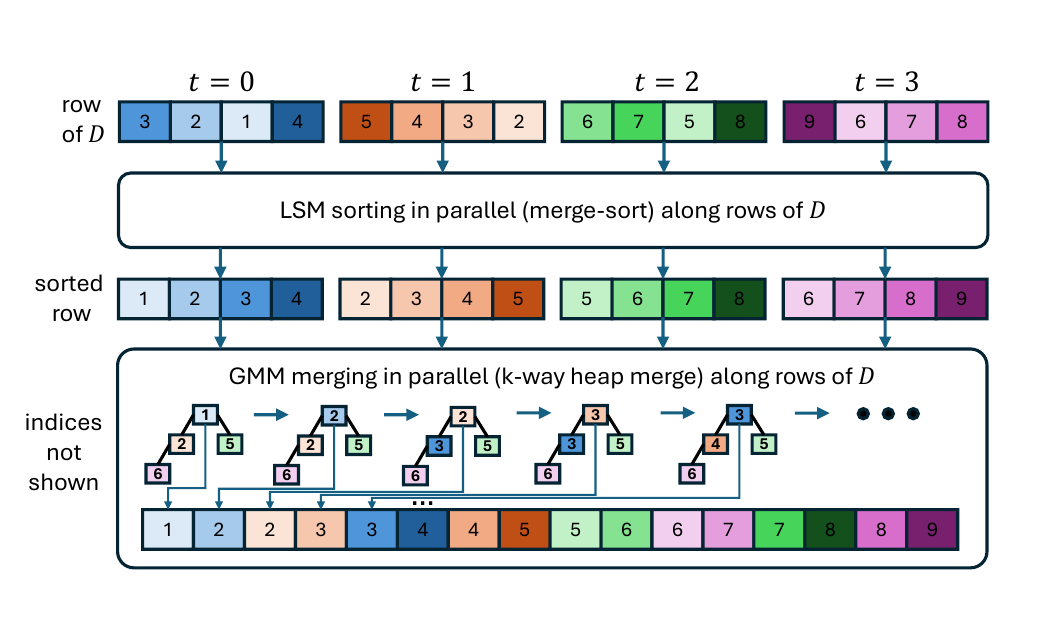}
\caption{Illustration of LSM and GMM operations, showing parallel index-based row-wise merge-sort in LSM and $k$-way heap merge in GMM to produce globally sorted neighbor lists with pipelined execution.}
\label{fig:lsm_gmm}}
\end{figure}

The Local Sorting Module (LSM), as defined in Algorithm \ref{alg:LSM} performs high-throughput sorting of the (distance, index) pairs produced by the DCM for each partition. Specialized sorting PEs, implemented as merge sort units, operate directly on the locally buffered outputs, producing partially sorted lists of candidate neighbors for each node within a partition. To support large-scale graphs, the LSM units are interconnected in a hierarchical fashion: the outputs of local sorts are routed to a global merge stage (see Fig. \ref{fig:lsm_gmm}). 
\begin{algorithm}
\footnotesize{
\begin{algorithmic}[1]
\caption{PE Allocation and Per-PE Row Sorting in the Local Sorting Module (LSM)}
\label{alg:LSM}
\Require Partitioned buffer with $n$ rows; number of parallel sorting PEs per partition: $P_{\text{sort}} \leq n$; number of top entries to keep: $k \leq m$
\Statex
\Comment{PE Allocation}
\For{$i = 0$ to $P_{\text{sort}} - 1$} \Comment{Parallel over rows}
    \State Allocate $\text{SortPE}_i$ for row $i$
\EndFor
\Statex
\Comment{Per-PE Row Sorting}
\For{$\text{SortPE}_i$ \textbf{in parallel}}
    \State $\text{row} \gets$ partition buffer $[i, 0\!:\!m-1]$
    \For{$j = 0$ to $k - 1$}
        \State $min\_idx \gets j$
        \For{$l = j+1$ to $m - 1$}
            \If{row$[l].$dist $<$ row$[min\_idx].\text{dist}$}
                \State $min\_idx \gets l$
            \EndIf
        \EndFor
        \State Swap $\text{row}[j] \leftrightarrow \text{row}[min\_idx]$
    \EndFor
    \State Output $(\text{dist}_{i,j}, i, j)$ to local sorted buffer
\EndFor
\end{algorithmic}}
\end{algorithm}

\subsubsection{Global Merging Module (GMM)}
The Global Merging Module (GMM), as described in Algorithm \ref{alg:GMM} merges locally sorted outputs from all partitions into a globally sorted list of neighbor candidates for each node. This module ensures the top-$k$ or dilated neighbor selection operates on the complete set of candidate distances. The GMM operates by receiving streams of (distance, index) pairs from the LSM, using a $k$-way merge based on a distributed min-heap. At initialization, the first element from each locally sorted stream is inserted into the min-heap. The GMM then extracts the minimum element from the heap, guaranteed to be the smallest remaining distance, and outputs it to the globally merged list. After each extraction, the next element from the corresponding stream is inserted into the heap, and the heap property is restored. This process continues until all streams are exhausted, yielding a single, fully sorted list of (distance, index) pairs for each node.
\begin{algorithm}
\footnotesize{
\begin{algorithmic}[1]
\caption{Global Merging Module (GMM): $k$-way Merge of Sorted Streams}
\label{alg:GMM}
\Require $Q$ sorted input streams, each with (distance, index) pairs for a node; number of streams to merge: $Q$; number of neighbors to select: $k$
\Statex
Initialize min-heap for $Q$ streams
\State Create min-heap $H$ of size $Q$
\For{$q = 0$ to $Q - 1$}
    \If{stream $q$ is not empty}
        \State Insert $(\text{stream}_q[0], q, 0)$ into $H$
    \EndIf
\EndFor
\Statex{$k$-way merge to produce global top-$k$}
\State $merged \gets$ empty list
\While{$|merged| < k*d$ \textbf{and} $H$ is not empty}
    \State $(pair, q, idx) \gets$ extract-min from $H$
    \State Append $pair$ to $merged$
    \If{stream $q$ has more elements}
        \State Insert $(\text{stream}_q[idx+1], q, idx+1)$ into $H$
    \EndIf
\EndWhile
\State Output $merged$ as globally sorted top-$k$ list
\end{algorithmic}}
\end{algorithm}

\subsubsection{Neighbor Selection Module}
The Neighbor Selection Module is implemented as a row of PEs which facilitate parallel selection of graph neighbors according to the dilated k-nearest neighbor (KNN) strategy. Each PE is assigned a segment of the globally sorted (distance, index) list for a node and is responsible for extracting the indices of neighbors at regular strides, as specified by the dilation factor $d$. The PEs operate across multiple nodes and candidate lists, with the number of parallel units parameterized over $p_{col}$ to match the desired throughput and available FPGA resources. Within each PE, a state machine iterates through its assigned sorted list, selecting every $d$-th entry from the top $k \times d$ candidates and writing the selected indices into the OB. On-chip BRAM is used to buffer both the input candidate lists and the output indices, ensuring low-latency data access and efficient pipelined operation. The modular structure of the neighbor selection PEs allows the accelerator to scale to larger graphs and higher degrees of parallelism.

We summarize cycle estimates for each accelerator module in Table~\ref{table:cycles}, based on ViG-Tiny (14$\times$14 patches, $N=M=196$, $D=192$, $k=8$, $d=2$, $m=28$, $P_{\text{row}}=P_{\text{col}}=14$, $P_{\text{vec}}=8$, $P_{\text{sort}}=7$, $Q=7$). DDR4 traffic ranges from 500 KB (Tiny) to over 50 MB for larger models, reflecting scalability with resolution and model size. Note distances are in 32-bit floats and indices in 16 bit unsigned integer.

\begin{table}
\centering
\footnotesize{\caption{Cycle estimates for each module as per our performance model.}
\label{table:cycles}}
\resizebox{0.3\textwidth}{!}{
\renewcommand{\arraystretch}{1.5}
\begin{tabular}{lcc}
\toprule
\textbf{Module} & \textbf{Cycles} & \textbf{Formula} \\
\midrule
DCM & 4,704 & $\left\lceil \dfrac{N}{P_\text{row}} \right\rceil \cdot \left\lceil \dfrac{M}{P_\text{col}} \right\rceil \cdot \left\lceil \dfrac{D}{P_\text{vec}} \right\rceil$ \\
LSM & 3,920 & $\left\lceil \dfrac{N}{P_\text{sort}} \right\rceil \cdot \left[m \cdot \left\lceil \log_2 m \right\rceil\right]$ \\
GMM & 4,704 & $N \cdot k \cdot \left\lceil \log_2 Q \right\rceil$ \\
NSM & 224   & $\left\lceil \dfrac{N}{Q} \right\rceil \cdot k$ \\
\bottomrule
\end{tabular}%
}
\end{table}

%% file: 4-exp.tex
\section{Experiments}
\label{sec: exp}


\begin{figure*}
  \centering
  \begin{subfigure}[t]{0.48\textwidth}
    \centering
    \includegraphics[width=\textwidth]{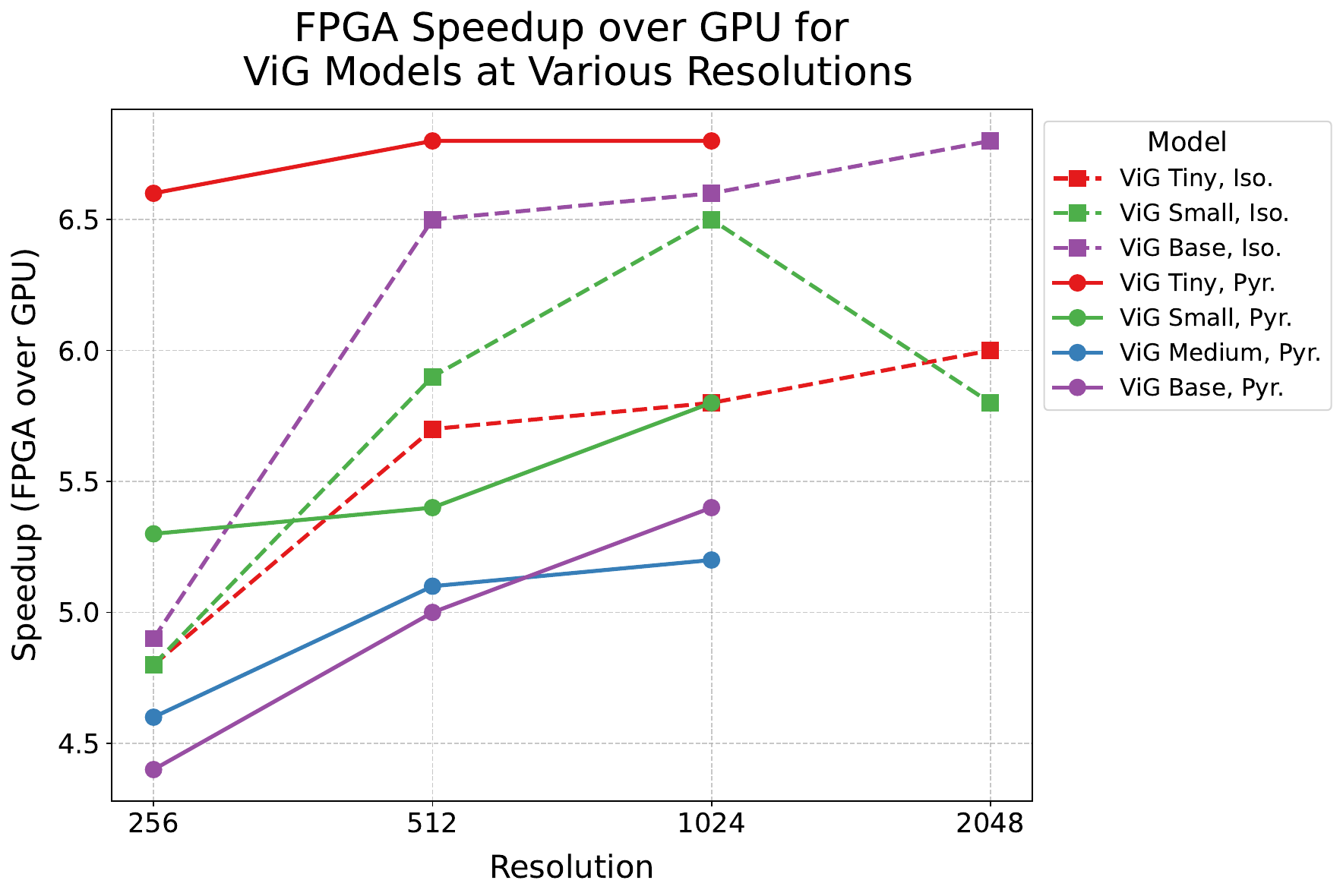}
    \caption{Speedup over GPU for ViG models (Ti: Tiny, S: Small, M: Medium, B: Base) and architectures (Pyr.: Pyramidal, Iso.: Isotropic) across input resolutions. Lines not extending towards 2048 are due to GPU memory limitations.}
    \label{fig:figSpeedGPU}
  \end{subfigure}
  \hfill
  \begin{subfigure}[t]{0.48\textwidth}
    \centering
    \includegraphics[width=\textwidth]{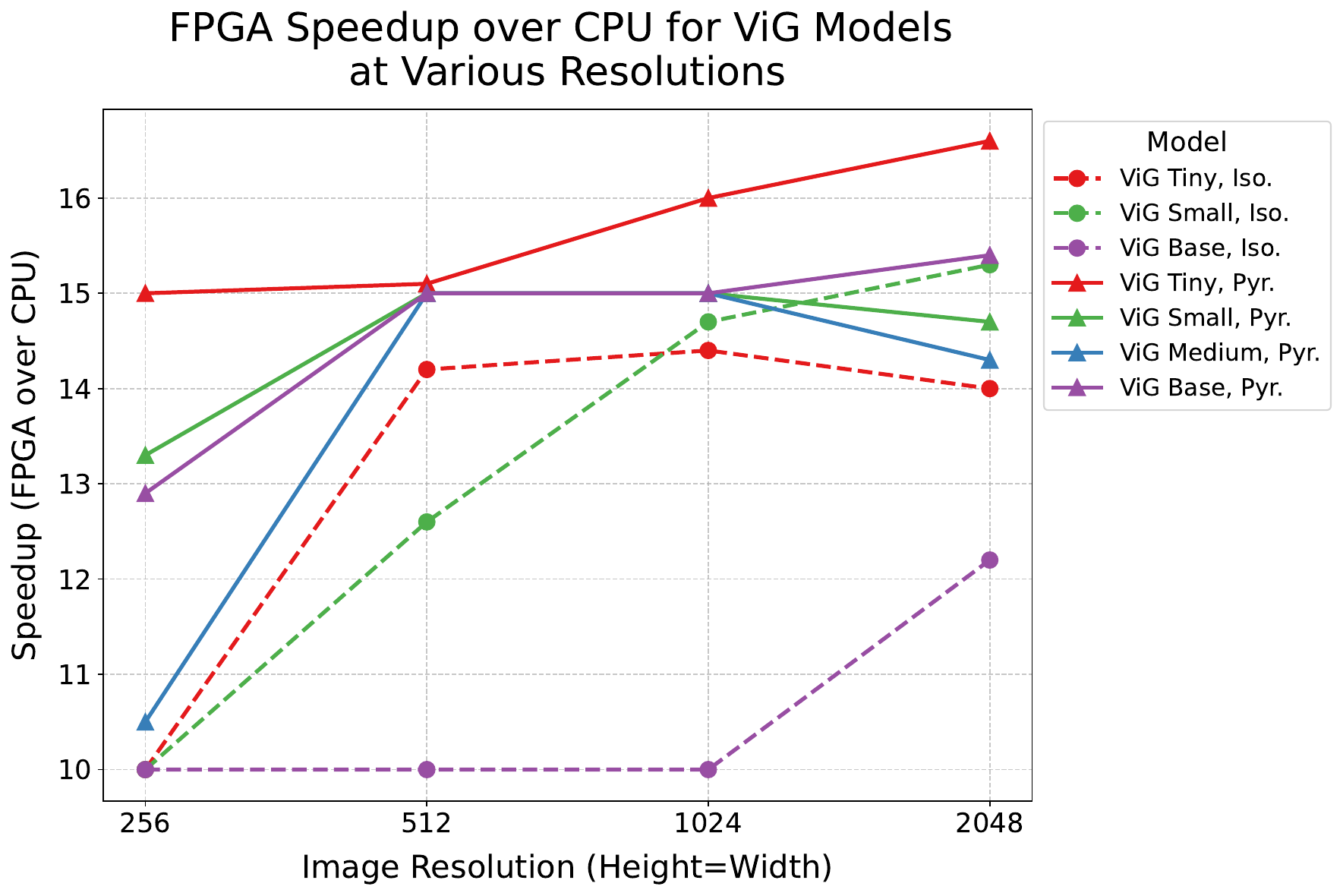}
    \caption{Speedup over CPU, with details matching (a).}
    \label{fig:figSpeedCPU}
  \end{subfigure}
  \caption{End-to-end speedup of our FPGA accelerator over GPU and CPU baselines across ViG models and image resolutions.}
  \label{fig:speedup_profiles}
\end{figure*}

\subsection{Hardware Implementation}

The FPGA accelerator is implemented on the Xilinx Alveo U280 platform, achieving a 600 MHz clock rate as reported by post place-and-route analysis using Vitis HLS 2022.2. This achieved timing is enabled by a deeply pipelined architecture and static parallelism configuration, which together minimize critical path delays across the design. Our resource utilization is depicted in Table \ref{tab:fpga_resources}. The partition size is set to $m = 28$, with a 2D PE mesh of $P_{\text{row}} = P_{\text{col}} = 8$, vector-level parallelism $P_{\text{vec}} = 8$, sorting parallelism $P_{\text{sort}} = 8$, and $Q = 8$ total partitions. The partition size $m=28$ fits BRAM limits while enabling efficient merge-sort operations. The PE mesh dimensions match available DSP and memory resources, supporting concurrent processing of node and co-node pairs without routing congestion. Vector parallelism ($P_{\text{vec}} = 8$) leverages SIMD-style computation in the feature domain, while $P_{\text{sort}} = 8$ and $Q = 8$ maximize throughput in neighbor selection by enabling concurrent partition processing.

\begin{table}
\centering
\footnotesize{
\caption{Resource utilization for our accelerator compared to available resources on Alveo U280.}
\label{tab:fpga_resources}}
\resizebox{0.4\textwidth}{!}{
\begin{tabular}{lcccc}
\toprule
\textbf{Resource} & \textbf{DSP} & \textbf{LUT} & \textbf{BRAM} & \textbf{URAM} \\
\midrule
Used              & 25           & 27,822       & 148           & N/A \\
Available (U280)  & 9,024        & 1,704,448    & 2,160         & 960 \\
\bottomrule

\end{tabular}%
}
\end{table}

\subsection{Baseline Platforms and Experimental Setup}
\label{sec:baselines}

All experiments are run on a dual-socket AMD EPYC 7763 server with 128 cores and 256 threads across 16 NUMA nodes using 128 threads and 64 cores, alongside GPU baselines on an NVIDIA RTX A5000 (24 GB GDDR6, 8192 CUDA cores) with 81\% utilization. Together with the Xilinx Alveo U280 FPGA, these platforms hardware from the same recent technology generation, ensuring a fair comparison. We evaluate isotropic and pyramidal ViG variants using the official PyTorch codebase ~\cite{VisionGNNGithub,VisionGNN}. We benchmark the runtime of DIGC for a single image for resolutions from $256 \times 256$ to $2048 \times 2048$, timing the Alveo U280's performance using the Vivado Implementation Timing Reports and Static Timing Analysis (STA) and assuming that the feature vectors are stored in memory prior to all measurements.

\subsection{Results}

\begin{table}
\centering
\caption{Runtime (ms) of DIGC for one image for ViG models across CPU, GPU, and FPGA platforms at various resolutions. CPU and GPU are followed by speedup achieved on FPGA.}
\label{tab:runtime}
\resizebox{0.4\textwidth}{!}{
\begin{tabular}{lccc c}
\toprule
\textbf{Model} & \textbf{Resolution} & \textbf{CPU Runtime (Speedup)} & \textbf{GPU Runtime (Speedup)} & \textbf{FPGA Runtime} \\
\midrule

\multirow{4}{*}{ViG Tiny, Isotropic}
 & 256$\times$256   & 15.625 (10.0$\times$)      & 8.567 (4.8$\times$)   & 1.762 \\
 & 512$\times$512   & 236.959 (14.2$\times$)     & 95.373 (5.7$\times$)  & 16.696 \\
 & 1024$\times$1024 & 657.283 (14.4$\times$)     & 265.704 (5.8$\times$) & 45.728 \\
 & 2048$\times$2048 & 1049.967 (14.0$\times$)    & 449.982 (6.0$\times$) & 74.997 \\
\midrule

\multirow{4}{*}{ViG Small, Isotropic}
 & 256$\times$256   & 20.231 (10.0$\times$)      & 9.619 (4.8$\times$)   & 2.023 \\
 & 512$\times$512   & 429.911 (12.6$\times$)     & 201.292 (5.9$\times$) & 33.991 \\
 & 1024$\times$1024 & 621.507 (14.7$\times$)     & 274.405 (6.5$\times$) & 42.151 \\
 & 2048$\times$2048 & 1371.507 (15.3$\times$)    & 608.9468 (6.8$\times$) & 89.551 \\
\midrule

\multirow{4}{*}{ViG Base, Isotropic}
 & 256$\times$256   & 31.748 (10.1$\times$)      & 15.698 (4.9$\times$)  & 3.175 \\
 & 512$\times$512   & 522.291 (10.0$\times$)     & 337.478 (6.5$\times$) & 52.229 \\
 & 1024$\times$1024 & 832.067 (10.0$\times$)     & 555.486 (6.6$\times$) & 83.207 \\
 & 2048$\times$2048 & 1394.358 (12.2$\times$)    & 116.169 (6.7$\times$) & 114.436 \\
\midrule

\multirow{4}{*}{ViG Tiny, Pyramid}
 & 256$\times$256   & 15.124 (15.0$\times$)      & 6.688 (6.6$\times$)   & 1.008 \\
 & 512$\times$512   & 61.271 (15.1$\times$)     & 27.709 (6.8$\times$)  & 4.031 \\
 & 1024$\times$1024 &257.968 (16.0$\times$)     & 109.165 (6.8$\times$)  & 16.123 \\
 & 2048$\times$2048 & 904.5008  (16.6$\times$)    & N/A                   & 54.488 \\
\midrule

\multirow{4}{*}{ViG Small, Pyramid}
 & 256$\times$256   & 27.266 (13.3$\times$)      & 10.847 (5.3$\times$)  & 2.051 \\
 & 512$\times$512   & 69.015 (15.0$\times$)      & 25.025 (5.4$\times$)  & 4.601 \\
 & 1024$\times$1024 & 276.013 (15.0$\times$)     & 107.343 (5.8$\times$)  & 18.401 \\
 & 2048$\times$2048 & 1084.003 (14.7$\times$)    & N/A                   & 73.600 \\
\midrule

\multirow{4}{*}{ViG Medium, Pyramid}
 & 256$\times$256   & 23.118 (10.5$\times$)      & 10.265 (4.6$\times$)  & 2.208 \\
 & 512$\times$512   & 72.419 (15.0$\times$)      & 24.675 (5.1$\times$)  & 4.828 \\
 & 1024$\times$1024 & 289.625 (15.0$\times$)     & 100.894 (5.2$\times$)  & 19.308 \\
 & 2048$\times$2048 & 1205.447 (14.3$\times$)    & N/A                   & 84.230 \\
\midrule

\multirow{4}{*}{ViG Base, Pyramid}
 & 256$\times$256   & 30.187 (12.9$\times$)      & 10.358 (4.4$\times$)  & 2.346 \\
 & 512$\times$512   & 80.678 (15.0$\times$)      & 27.082 (5.0$\times$)  & 5.379 \\
 & 1024$\times$1024 & 322.641 (15.0$\times$)     & 116.419 (5.4$\times$) & 21.509 \\
 & 2048$\times$2048 & 1790.493 (15.4$\times$)    & N/A                   & 116.033 \\
\bottomrule
\end{tabular}}
\end{table}

\begin{table}
\centering
\caption{Overall speedup (\(\times\)) from offloading graph construction to FPGA for ViG models (256$\times$256 resolution).}
\label{tab:speedup}
\resizebox{0.3\textwidth}{!}{
\begin{tabular}{lcc}
\toprule
Model & CPU + FPGA & GPU + FPGA \\
\midrule
Ti, Iso. & 2.13$\times$ & 1.85$\times$ \\
S, Iso. & 4.04$\times$ & 1.85$\times$ \\
B, Iso. & 4.62$\times$ & 2.10$\times$ \\
Ti, Pyr. & 2.41$\times$ & 1.83$\times$ \\
S, Pyr. & 2.68$\times$ & 1.98$\times$ \\
M, Pyr. & 3.60$\times$ & 2.00$\times$ \\
B, Pyr. & 4.04$\times$ & 1.91$\times$ \\
\bottomrule
\end{tabular}
}
\end{table}

Table~\ref{tab:runtime} presents the runtime and speedup of our FPGA accelerator on DIGC for a singular image compared to CPU and GPU baselines across increasing image resolutions, highlighting its ability to maintain high throughput where GPU baselines fail due to memory exhaustion. This is enabled by a hierarchical parallelism strategy, partitioned BRAM assignments that eliminate contention, and a fully streaming pipeline sustains dataflow across all stages. We achieve end-to-end inference speedup by offloading the DIGC phase of running end-to-end inference for ViG models to the FPGA.  As shown in Table~\ref{tab:speedup}, CPU + FPGA configurations yield a $2.13\times$-$4.62\times$ speedup across ViG variants at $256 \times 256$ resolution, while GPU + FPGA setups achieve $1.83\times$-$2.10\times$ gains. The improvement is most pronounced for isotropic and larger models, where DIGC accounts for a significant share of inference latency. We present plots illustrating the scalability of our FPGA accelerator in Figures \ref{fig:figSpeedCPU} and \ref{fig:figSpeedGPU} across varying input resolutions. These results highlight consistent latency and speedup gains over both CPU and GPU baselines as image size increases. This confirms that the proposed design overcomes scalability limits of GPU implementations and delivers robust performance for high-resolution ViG workloads.

%% file: 5-concl.tex
\section{Conclusion \& Future Work}
Our accelerator demonstrates substantial performance improvements, achieving a 10.1× to 16.6× speedup in dynamic image graph construction (DIGC) over CPU and a 4.4× to 6.8× speedup over GPU. Integrated into end-to-end ViG inference, it delivers a 2.13× to 4.62× speedup versus CPU + FPGA and 1.83× to 2.10× versus GPU + FPGA configurations. The architecture scales with input resolution, offering greater benefits for high-resolution workloads. 

Due to the modular nature of our accelerator, the graph construction approach can be generalized by adjusting the mechanism used to compute similarity between image patches. This design choice supports diverse graph construction strategies, such as clustering-based approaches exemplified by ClusterViG and greedy edge-selection techniques used in GreedyViG. The ability to integrate alternative construction schemes without altering the core architecture enhances the applicability of our accelerator for varying vision applications.